# Software Engineering Education by Example

**Nacer Boudjlida, Jean-Pierre Jacquot, Pascal Urso**

*Nancy Université, UHP Nancy 1, LORIA, UMR 7503*
*Campus Scientifique, BP 70239*
*54506 Vandoeuvre Lès Nancy CEDEX, France*
*{Nacer.Boudjlida, Jean-Pierre.Jacquot, Pascal.Urso}@loria.fr*

ABSTRACT: *Based on the old but famous distinction between"in the small" and "in the large" software development, at Nancy Université, UHP Nancy 1, we experience for a while software engineering education thanks to actual project engineering. This education method has the merit to enable students to discover and to overcome actual problems when faced to a large project which may be conducted by a large development team. The mode of education is a simulation of an actual software engineering project as encountered in ``real life'' activities.*

## 1. Introduction

In this paper we claim that the education of software engineering processes requires "non classical" education methods. Indeed, engineering of *in the large* software reveals many non technical problems for which usual lectures are not adequate. This *"in the large"* dimension enables students to scale-up in software engineering. Our claim is that one of the success factors is to put students in a real (or at least a realistic) large software development situation and to accompany them by complementary professional lectures, like project management, quality assessment and so on.

We report hereafter on the education process of this discipline as applied for more than 15 years at Nancy University, University Henri Poincaré Nancy 1, at the second level of a master degree in Computer Science (during the 3[rd] semester of the master) dedicated to software engineering.

## 2. Pragmatic software engineering education: pre-requisite

The approach for educating in engineering of large software has as prerequisites, scientific and technologic software basis that are supposed to be acquired and experienced. These include programming and design fundamentals: software systems design methods, programming paradigms and programming languages, software systems architectures (like client/server architectures, distributed applications, Web-based applications, etc.). Moreover, fundamentals properties of programs are also supposed to be mastered: they include formal properties of software programs, like correctness, completeness, termination and complexity. Usually, these skills and knowledge are experimented against small-size projects. But, going toward large-size projects, i.e. projects whose development requires more than one or two persons, additional knowledge and skills are mandatory. In this education program, these additions are twofold:

**1.** *Computer science and technology lectures:* they aim at fitting students with state of the art and state of the practice in software engineering. This track is a completion of the academic curricula in computer science and it includes lectures and practice in emergent technologies. Currently they encompass component-based programming, service oriented architectures, model driven architectures and model-driven engineering.

**2.** *Project management accompanying lectures:* during the actual project development additional lectures are provided to fit students with skills concerning project management, project planning, human resource management, enterprise organization and management and so on. These non technologic lectures are supposed to bring to students professional competencies.

## 3. Pragmatic software engineering education: organization and roles

Considering that lecturing of this discipline is boring for the lecturer as well as for the audience, its education is done thanks to practice. Indeed, a rough definition of a project, usually provided by an industrial partner, is given to the students. The project team is composed with all the registered students (the size of the project team usually ranges from 15 to 20 members). An initial organization of the team into groups is proposed to the students. This organization emphasizes two roles, each of them being played by one student: (i) *project manager* and (ii) *project administrator*. Additional suggested *project groups* concern requirement, documentation, software specification, design and development, quality insurance and software testing. However, the project team and especially the project manager is free to re-organize the groups and the task assignments during the actual project development. Further, the *follow-up* of the project is performed by two professors who periodically meet the project manager and the project administrator, the role of the industrial partner being a project *customer role*.

One of the main recommendation that is made to the students is to be autonomous, to take initiatives and to be responsible of their job and of the associated logistics. Indeed, as computer resources, the students are provided with their proper set of computers and network: They can install (and uninstall), configure and exploit any technology that may be required by the project (like version management systems, development frameworks, plug-ins, shared repositories and so on). From the infrastructure quality insurance, the deontology and the legal aspects, the whole exploitation of the infrastructure is under the collective responsibility of the whole project team.

**4. Pragmatic software engineering education: academic evaluation of the result**

The ultimate aim of the education being giving grades to students, the evaluation of this discipline is twofold (i) there is an "academic" evaluation which is provided by the industrial project partner and the professors who performed the follow-up of the project execution and (ii) every team member is invited to evaluate the other team members and himself from different perspectives (technical skills, personal involvement in the project, initiatives, communication skills, and so on). Both the evaluations are then combined to get a unique and final grade.

**5. Pragmatic software engineering education: learned lessons**

Learned lessons are examined hereafter from various aspects:
- This type of education is experienced for a while with a very good feedback from the students themselves, especially when they defend the industrial internship work they have to perform during, at least, 4,5 months in an industrial context. Even the industrial internship hosting partner provides us with a good feedback concerning the abilities of our students in participating in large-scale projects.
- Basing the education on a real problem to solve and not, as usual, on a "toy problem", is very motivating for the students.
- In the early phases of the project, students fears responsibilities and autonomy but they usually get very quickly used to them. For example, no notable incident occurred regarding the quality insurance, the deontology and the legal aspects relative to the exploitation of the computer infrastructure that is provided to the students.
- By the end of the project, the students are aware of non technologic but very important success factors like resource management, planning and the quality of the human relationships among the project team's members.
- Considering the self-evaluation of the results, students are usually reluctant to evaluate their "friends" but we observed that they are very objective in their evaluation since these evaluations very often meet the opinion the accompanying professors and the industrial partner have on every student.

▪ Despite the fact that some project results did not meet the industrial partner expectations, the overall benefits for the students is valuable.

▪ An observed important success factor is the existence of one or more students fitted with a leadership spirit. Usually this type of character, not necessarily coming from the student who plays the project manager role,   deeply influences at the same time the quality of the results, the project team cohesion and the atmosphere inside the project team. Unfortunately, this type of personality is not revealed every year.

We are intimately convinced that in the large software engineering education in a "non classical" way brings more knowledge, skills and experience than classical lectures in the field. However, educating by example requires more effort from the students    than a traditional education of this discipline. This effort is especially needed in the early phase of the guiding project example when they are invited to perform a supervised self-learning of complementary subjects like quality insurance, project management and so on. But a remarkable benefit is that students are more or less obliged to become autonomous very rapidly and therefore to have a professional behaviour rather than passive student behaviour.